\documentclass[aps,prd,preprint,floatfix]{revtex4}
\usepackage{epsfig}

\begin{document}

\preprint{\rightline{ANL-HEP-PR-02-091}}

\title{The $N_t=4$ finite temperature phase transition for lattice QCD with
a weak chiral 4-fermion interaction.}

\author{J.~B.~Kogut}
\address{Dept. of Physics, University of Illinois, 1110 West Green Street,
Urbana, IL 61801-3080, USA}
\author{D.~K.~Sinclair}
\address{HEP Division, Argonne National Laboratory, 9700 South Cass Avenue,
Argonne, IL 60439, USA}

\begin{abstract}
We study the finite temperature phase transition of lattice QCD with an
irrelevant chiral 4-fermion interaction and two massless quark flavours, on
$8^3 \times 4$ and $12^2 \times 24 \times 4$ lattices. The strength of the
4-fermion interaction was reduced to half the minimum value used in previous 
simulations, to study how the nature of this phase transition depends on this
additional interaction. We find that the transition remains first order as
the 4-fermion coupling is reduced. Extending our earlier studies indicates 
that for sufficiently large 4-fermion coupling, the transition is probably
second order.
\end{abstract}
\maketitle

\pagestyle{plain}
\parskip 5pt
\parindent 0.5in

\section{Introduction}

In a recent series of papers \cite{kls,ks}, we have used a new staggered
fermion action modified to include an irrelevant 4-fermion interaction which
preserves the flavour symmetries of the standard staggered action, to study
the chiral symmetry restoring finite temperature phase transition of QCD. (For
earlier work by others proposing similar actions see \cite{kmy,brower}.) In
particular, we study the transition with two massless quark flavours. Such
simulations are impossible with the standard action, since the Dirac operator
becomes singular at zero quark mass. On lattices with temporal extent $N_t=4$,
the transition appeared to be first order \cite{kls}, while at $N_t=6$ it
appeared to be second order as expected \cite{pw}, but with the critical
indices of a tricritical point \cite{ks}, rather than the $O(4)/O(2)$ indices
expected from universality arguments. We interpreted these results to indicate
that at these large lattice spacings $a$, the additional interactions due to
discretization, which vanish as $a^2$, are large enough to affect the nature
of the transition. For this reason, we are currently performing simulations at
$N_t=8$. One such interaction is our additional 4-fermion interaction, so we
need to vary its coefficient in an attempt to determine how important it is in
determining the nature of the phase transition. At $N_t=6$ we ran at 2
different (small) values of this coupling and concluded that although the
$\beta_c=6/g_c^2$ of the transition changed as this coupling was varied, its
nature did not. In this paper we present the results of studies of the
dependence of the $N_t=4$ transition on the 4-fermion coupling.

The action we use is the staggered lattice version of the continuum Euclidean
Lagrangian density,
\begin{equation}
{\cal L}=\frac{1}{4}F_{\mu\nu}F_{\mu\nu}
        +\bar{\psi}(D\!\!\!\!/+m)\psi
        -{\lambda^2 \over 6 N_f}[(\bar{\psi}\psi)^2
                          -(\bar{\psi}\gamma_5\tau_3\psi)^2].
\label{eqn:lagrangian}
\end{equation}
For details of the staggered lattice transcription of this Lagrangian, using
auxiliary fields to render it quadratic in the fermion fields, thus allowing
simulations, we refer the reader to our earlier papers. The largest value of
$\gamma=3/\lambda^2$ used in our previous simulations at $N_t=4$, and the only
value for which we determined the nature of the transition, was $\gamma=10$.
We present the results of new simulations at $\gamma=20$ on $8^3 \times 4$ and
$12^2 \times 24 \times 4$ lattices, which indicate that the transition remains
first order. We have also extended the studies of the $\gamma=5$ transition
reported in our earlier work. These new studies show no sign of the
meta-stability observed at $\gamma=10,20$, suggesting that the $\gamma=5$
transition is second order.

In section~2 we present the results of these simulations. Section~3 gives our
conclusions.

\section{Simulations at $N_t=4$, $N_f=2$ and $\gamma=20,5$.}

We performed simulations using the hybrid molecular dynamics method with noisy
fermions allowing us to tune to 2 flavours. We ran on $8^3 \times 4$ and
$12^2 \times 24 \times 4$ lattices to observe finite size effects. On both
lattices our molecular dynamics time increment was chosen to be $dt=0.05$, 
which appeared adequate. Our quark mass was set to zero and $\gamma=20$.

\begin{figure}[htb]
\epsfxsize=6in
\centerline{\epsffile{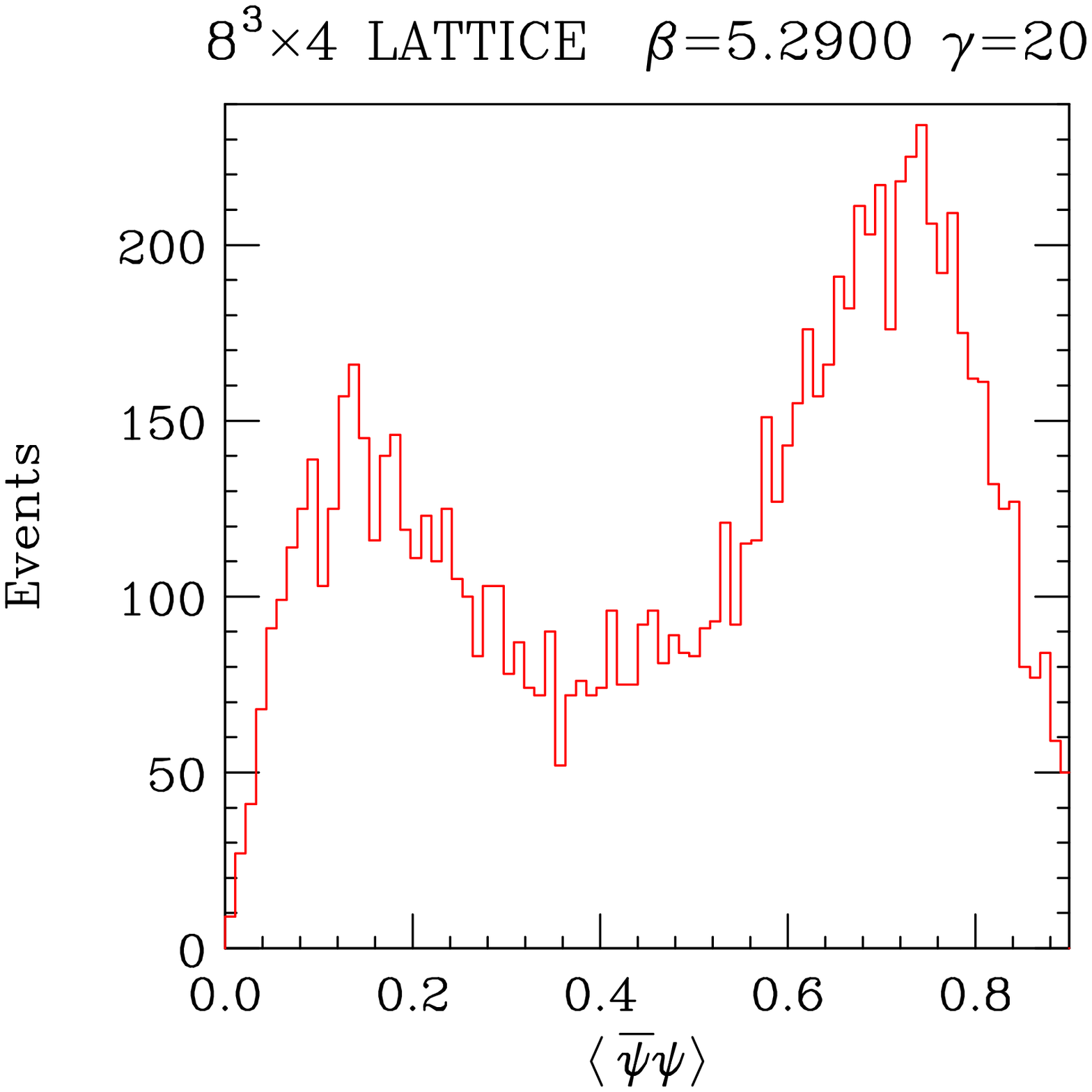}}
\caption{Histogram of $\langle\sqrt{\langle\bar{\psi}\psi\rangle^2-
\langle\bar{\psi}\gamma_5\xi_5\psi\rangle^2}\rangle$ measurements spaced at 2 
time-unit intervals on an $8^3 \times 4$ lattice at $\beta=5.29$.}
\label{fig:hist8}
\end{figure}

For the $8^3 \times 4$ simulations we ran for $20,000$ molecular dynamics time
units for $\beta=5.27$ and $7$ $\beta$ values in the range $5.285 \le \beta
\le 5.3$, and $10,000$ time units for 3 $\beta$s outside this range. We
observed evidence for metastability with 2-state signals both in the chiral
condensate and in the Wilson line (Polyakov loop) for $5.285 \lesssim \beta
\lesssim 5.3$, with the clearest signals at $\beta=5.29$ and $\beta=5.295$. A
histogram of the chiral condensate at $\beta=5.29$ is shown in
figure~\ref{fig:hist8}, showing two clearly separated peaks. From this we
would conclude that there is a first order transition. We estimate the
transition to occur at $\beta=\beta_c=5.2925(50)$.

\begin{figure}[htb]                                                           
\epsfxsize=6in                                                                
\centerline{\epsffile{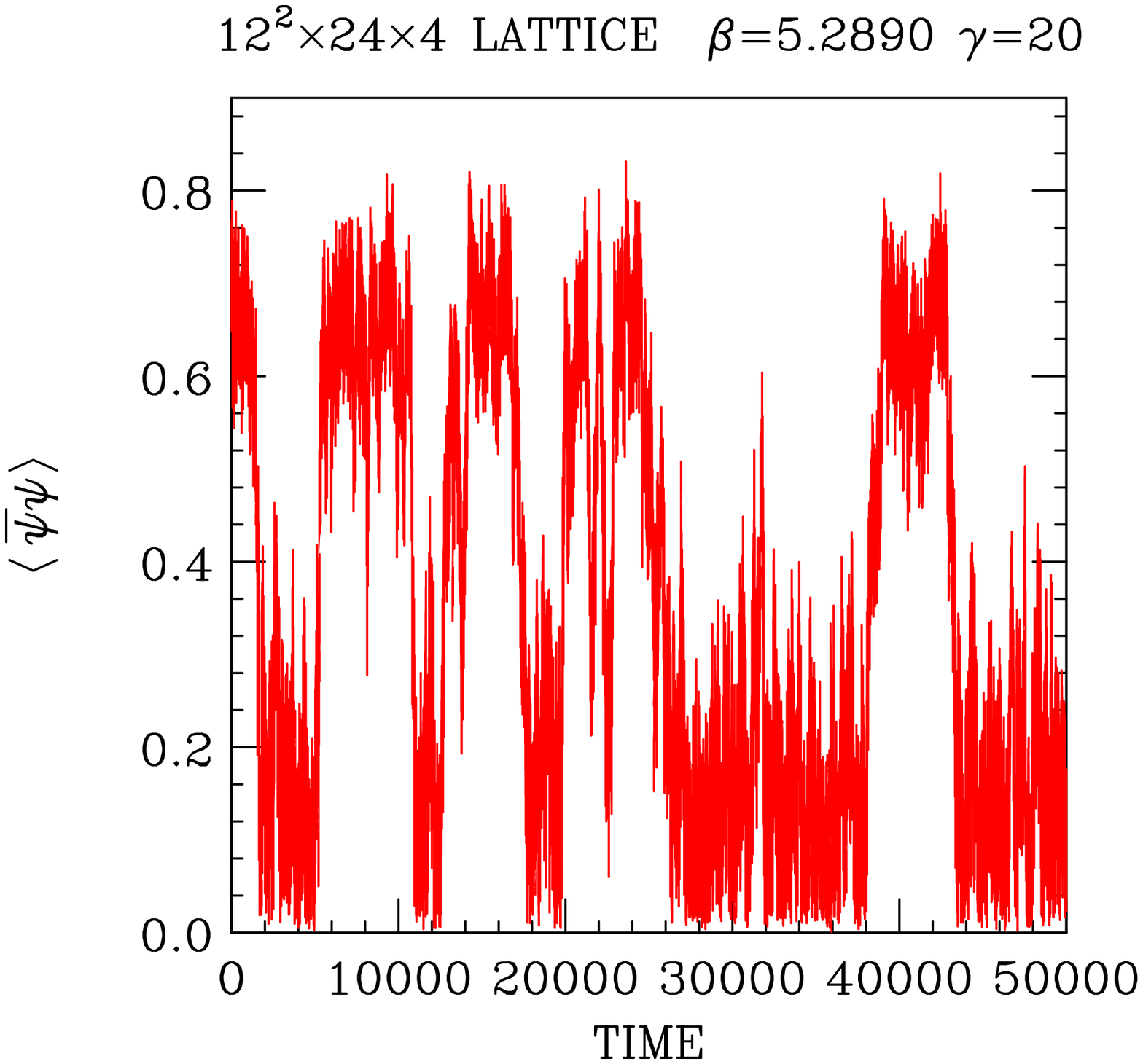}}
\caption{Time evolution of $\langle\sqrt{\langle\bar{\psi}\psi\rangle^2-
\langle\bar{\psi}\gamma_5\xi_5\psi\rangle^2}\rangle$ on a 
$12^2 \times 24 \times 4$ lattice at $\beta=5.289$.}
\label{fig:time}
\end{figure}

\begin{figure}[htb]                                                          
\epsfxsize=6in                                                               
\centerline{\epsffile{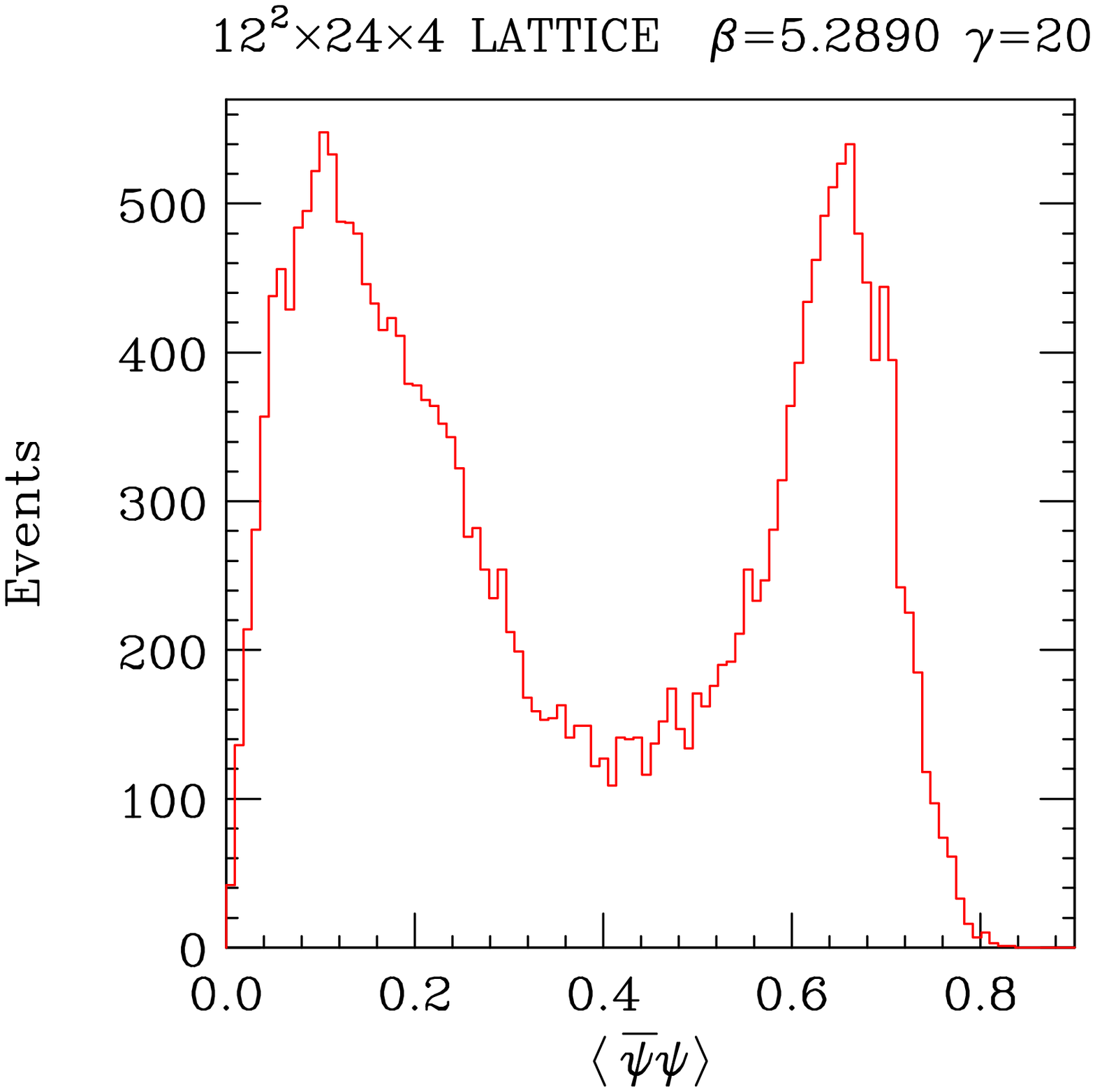}}                           
\caption{Histogram of $\langle\sqrt{\langle\bar{\psi}\psi\rangle^2-     
\langle\bar{\psi}\gamma_5\xi_5\psi\rangle^2}\rangle$ on a                    
$12^2 \times 24 \times 4$ lattice at $\beta=5.289$.}                         
\label{fig:hist12}                                                        
\end{figure}

For our $12^2 \times 24 \times 4$ simulations, we ran for $50,000$ time units
at $\beta=5.2875$, $\beta=5.289$ and $\beta=5.29$, close to the transition,
and for shorter `times' at 8 other $\beta$ values in the range 
$5.265 \le \beta \le 5.35$. We observed clear signals for metastability
with a well defined 2-state signal at $\beta=5.289$ and $\beta=5.29$, but not
outside this region, clear evidence that there is a first order transition at
$\beta_c=5.289(1)$. The time evolution of the chiral condensate at $\beta=5.289$
is shown in figure~\ref{fig:time}.
A histogram showing the two-state signal in the chiral condensate at 
$\beta=5.289$ is shown in figure~\ref{fig:hist12}.
The separation of the peaks in this histogram is $\approx 0.56$, while that
on the $\beta=5.29$ histogram on the $8^3 \times 4$ lattice is $\approx 0.61$.
The decrease in going to the larger lattice is small, leading us to the
conclusion that the first order transition is real, and not a small lattice
artifact. For the Wilson line, the peak separation is $\approx 0.22$ for the
larger lattice and $\approx 0.32$ for the smaller lattice. Most of this
change comes from a rise in the value of the Wilson line for the low temperature
phase, and might well be a fermion screening effect on the larger lattice.

\begin{figure}[htb]
\epsfxsize=6in
\centerline{\epsffile{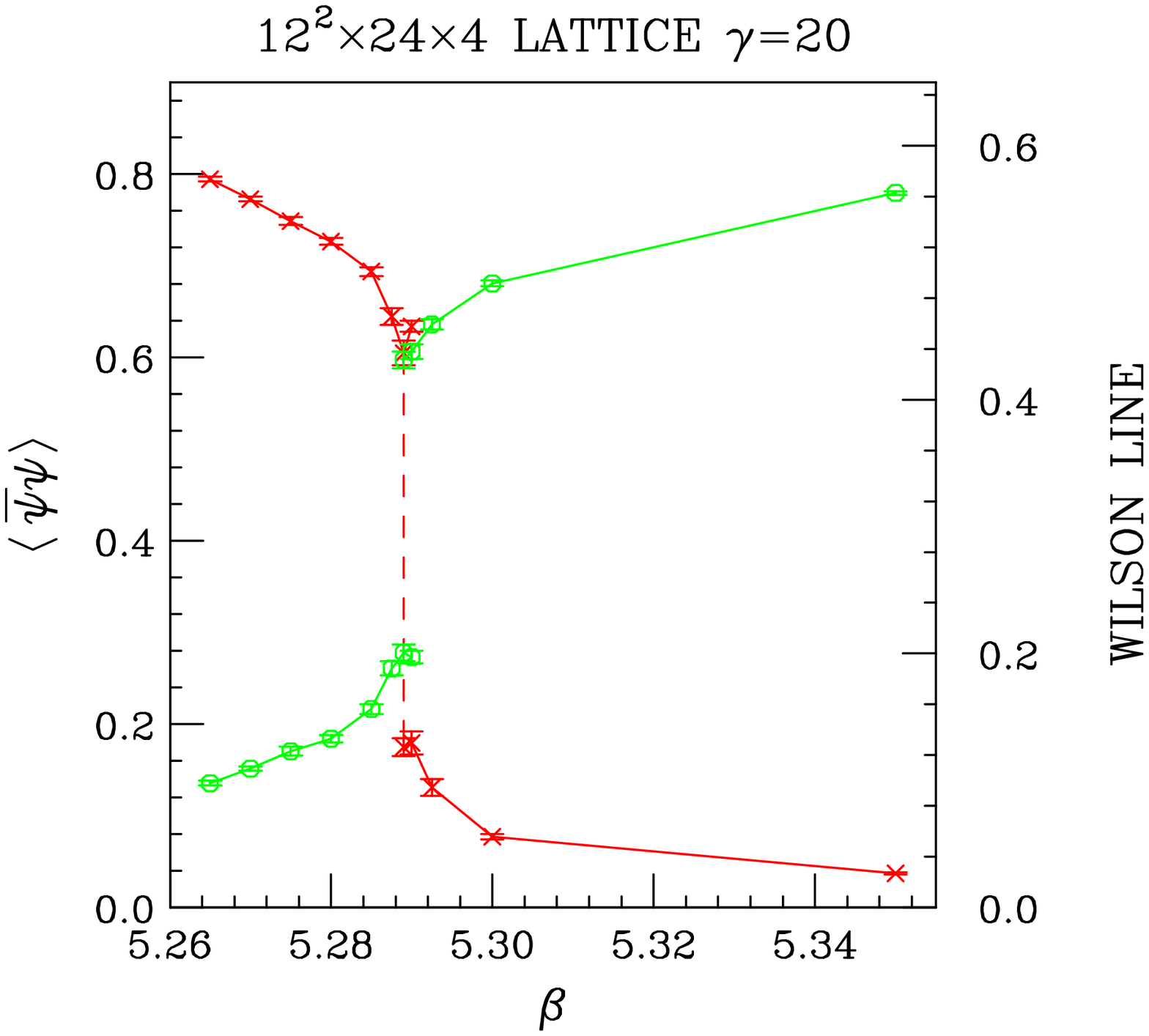}}
\caption{Chiral condensate and Wilson line on a $12^2 \times 24 \times 4$ 
lattice as functions of $\beta$ at $\gamma=20$.}
\label{fig:wil-psi}
\end{figure}

Finally, figure~\ref{fig:wil-psi} shows the two order parameters for the
$12^2 \times 24 \times 4$ lattice simulations as functions of $\beta$, showing
how abrupt the transition is.

\begin{figure}[htb]
\epsfxsize=6in               
\centerline{\epsffile{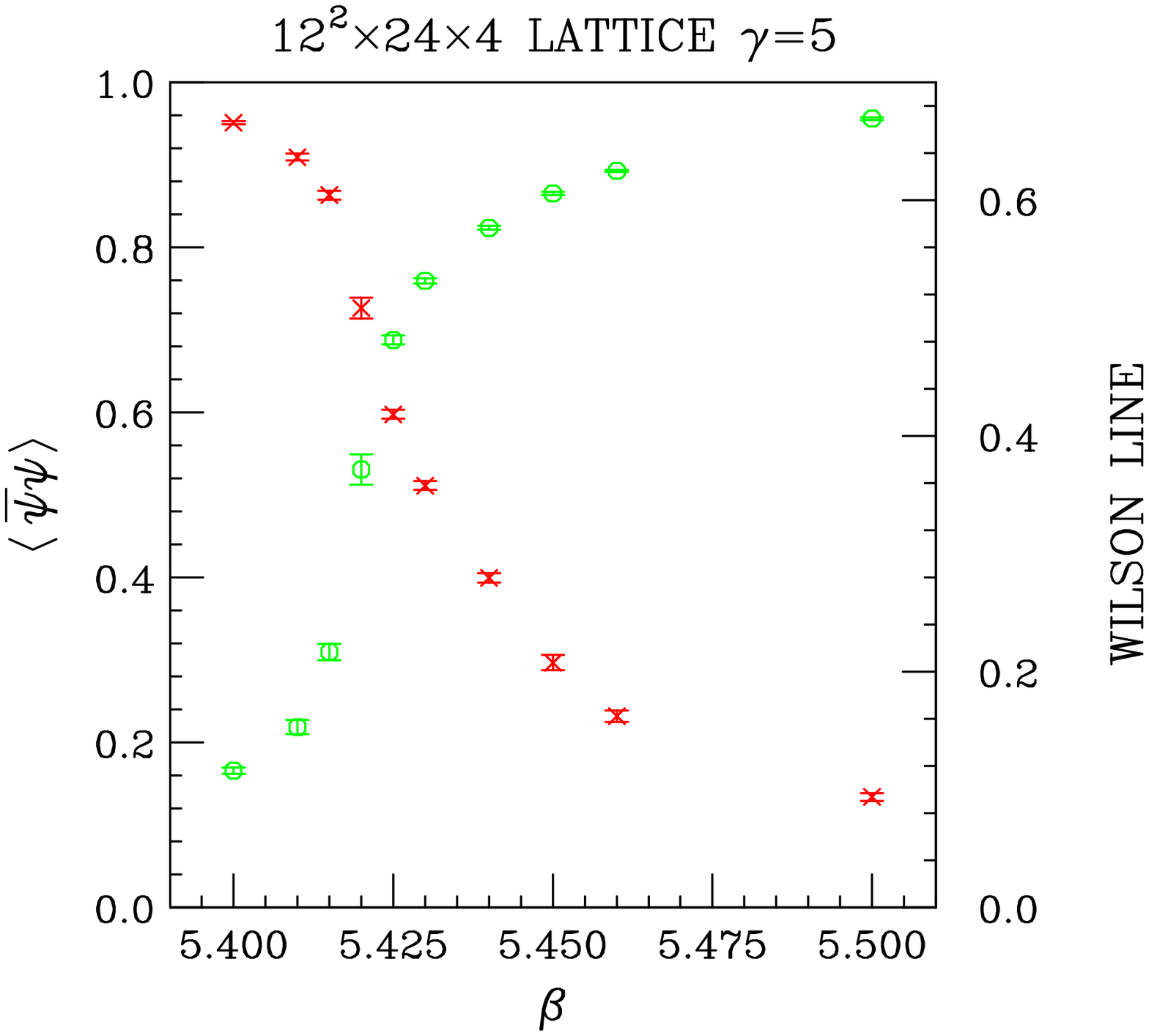}}
\caption{Chiral condensate and Wilson line on a $12^2 \times 24 \times 4$
lattice as functions of $\beta$ at $\gamma=5$.}                      
\label{fig:wil-psi5}                                                      
\end{figure}

Since our earlier studies at stronger 4-fermion couplings were inadequate to
determine the nature of the phase transition, we have extended the $\gamma=5$
simulations to $12^2 \times 24 \times 4$ lattices and added some $\beta$ values
in the crossover region to our $8^3 \times 4$ simulations. For the larger
lattice we ran for $50,000$ time units for each $\beta$ in the range $5.415
\le \beta \le 5.440$. Figure~\ref{fig:wil-psi5} shows the chiral condensate
and Wilson line for the $12^2 \times 24 \times 4$ lattice. The transition
appears smooth, and the time dependence of these order parameters shows no
sign of metastability for any of these $\beta$ values. This suggests that the
$\gamma=5$ chiral transition is second order and occurs at $\beta=5.425(5)$
\section{Conclusions}

We have simulated lattice QCD with 2 flavours of massless staggered quarks and
an additional chiral 4-fermion term on $N_t=4$ lattices. The chiral finite
temperature transition at weaker 4-fermion coupling ($\gamma=20$) appears to
be first order, as it was at stronger 4-fermion coupling ($\gamma=10$). The
chiral condensate in the confined phase at the transition was $0.60(1)$ at
$\gamma=20$, compared with $0.76(1)$ for $\gamma=10$ (in the deconfined phase
this condensate would vanish). The change in the Wilson line across the
transition is $0.23(1)$ for $\gamma=20$, compared with $0.33(1)$ at
$\gamma=10$. A priori, this softening of the transition as the 4-fermion
coupling weakens might suggest that it would become second order at vanishing
4-fermion coupling. However, we see that at stronger coupling, $\gamma=5$, the
transition is even softer and probably second order, so the situation is not
so simple.

We suggest that the order of the transition is determined primarily by the
`standard' ($\gamma$ independent) part of the action, and hence by $\beta$. 
On $N_t=4$ lattices, $\beta_c=5.289(1)$ at $\gamma=20$ compared with
$\beta_c=5.327(2)$ at $\gamma=10$ and $5.425(5)$ at $\gamma=5$. At 
$\gamma=10,20$ the lower values of $\beta_c$ mean larger lattice artifacts,
which we suggest make the transition first order. The larger $\beta_c$ at
$\gamma=5$, has smaller lattice artifacts, which we suggest leave it second
order. It is interesting to note that $\beta_c$ at $\gamma=5$ and $N_t=4$
is very close to $\beta_c$ at $\gamma=20$ for $N_t=6$, which we have determined
to be second order (tricritical), which is consistent with this scenario. The
reason the discontinuity in the $\gamma=10$ chiral condensate is greater than
that at $\gamma=20$ is that the stronger 4-fermion coupling enhances condensate
formation. The larger $\beta$ of the $\gamma=10$ transition increases the
Wilson line just above the transition over the corresponding $\gamma=20$ value.
Such a scenario suggests that at zero 4-fermion coupling $\gamma=\infty$, the
transition would remain first order. As was noted in reference~\cite{MILC},
the presence of this first order transition could explain the failure of
universal critical scaling at the $N_t=4$ finite temperature transition for
lattice QCD with 2 staggered fermion flavours as $m \rightarrow 0$
\cite{MILC,boyd,JLQCD,laermann}.

Finally, we should comment that, because this 4-fermion term also breaks the
same flavour symmetries as the standard staggered action, it has the potential
to significantly increase this breaking and make improving this action more
difficult. We are planning hadron spectroscopy calculations to determine how
bad this breaking might be. However, this additional ${\cal O}(a^2)$ symmetry
breaking could be reduced to ${\cal O}(a^4)$ by addition of the terms required
to make this 4-fermion interaction that of the staggered lattice
implementation of the $SU(4) \times SU(4)$ or 
$(SU(2) \times SU(2)) \times (SU(2) \times SU(2))$ Gross-Neveu model
\cite{gn}. The only problem is that such an action has a complex fermion
determinant, essential to the physics of interest. However, changing some of
the couplings from imaginary to real removes this difficulty, and it might be
possible to analytically continue results from real values to imaginary values
of these couplings, as has been suggested for similar terms in the highly
improved staggered actions proposed by the HPQCD collaboration \cite{HPQCD}.
We note that the 4-fermion term in our action, and at least some of the
additional 4-fermion terms  which we are suggesting, are present in the HPQCD
action after a Fierz transformation.

\section*{Acknowledgments}
DKS is supported by DOE contract W-31-109-ENG-38. JBK is supported in part by 
NSF grant NSF PHY-0102409. The $\gamma=20$ simulations were performed on the 
Cray J90's and Cray SV1's at NERSC. The $\gamma=5$ simulations were performed
on the HP Superdome at the University of Kentucky under an NRAC grant.


\begin{thebibliography}{999}
\bibitem{kls}
J.~B.~Kogut, J.-F.~Laga\"{e} and D.~K.~Sinclair, Phys. Rev. D58, 034504 (1998).
\bibitem{ks}
J.~B.~Kogut and D.~K.~Sinclair, Phys. Lett. B492, 228 (2000);
J.~B.~Kogut and D.~K.~Sinclair, Phys. Rev. D64, 034508 (2001).
\bibitem{kmy} 
K.~I.~Kondo, H.~Mino and K.~Yamawaki, Phys. Rev. D39, 2430 (1989).
\bibitem{brower} 
R.~C.~Brower, Y.~Shen and C.-I.~Tan, Boston University preprint BUHEP-94-3
(1994); R.~C.~Brower, K.~Orginos and C.-I.~Tan, Nucl. Phys. B(Proc. Suppl.) 42,
42 (1995).
\bibitem{pw}
R.~Pisarski and F.~Wilczek, Phys. Rev. D29, 338 (1984).
\bibitem{MILC}
C.~Bernard {et al.} (MILC collaboration) Nucl. Phys. B(Proc. Suppl.) 63A-C,
400 (1998); Phys. Rev. D61, 054503 (2000).
\bibitem{boyd}
G.~Boyd with F.~Karsch, E.~Laermann and M.~Oevers, Proceedings
of 10th International Conference {\it Problems of Quantum Field Theory},
Alushta, Crimea, Ukraine (1996).
\bibitem{JLQCD}
S.~Aoki, {\it et al.} (JLQCD Collaboration), Phys. Rev. D57, 3910 (1998).
\bibitem{laermann} 
E.~Laermann, Nucl. Phys. B(Proc. Suppl.) 63A-C, 114 (1998).
\bibitem{gn}
D.~J.~Gross and A.~Neveu, Phys. Rev. D10, 3235 (1974).
\bibitem{HPQCD}
Q.~Mason, {\it et al}, hep-lat/0209152 (2002).

\end{thebibliography}
\end{document}